\documentclass[twoside,a4paper]{article}
\usepackage[T2A]{fontenc}
\usepackage[cp1251]{inputenc}
\usepackage[english]{babel}
\usepackage{graphicx}
\newcommand{\beq}{\begin{equation}}
\newcommand{\eeq}{\end{equation}}
\newcommand{\bea}{\begin{eqnarray}}
\newcommand{\eea}{\end{eqnarray}}

\begin{document}

\vbox{}

\begin{center}
{\LARGE Revision of distance to SS\,433}
\vspace{0.7cm}

{\large  A A Panferov}

IPT, Togliatti State University, Russia

E-mail: panfS@yandex.ru
\end{center}
\vspace{0.7cm}

\begin{abstract}
Critical analysis shows that all estimates of the velocity of the radio jets of SS\,433 and the distance to the object based on the relativistic effect of light travel time (the light travel effect) are not accurate enough to be conclusive.
From our elaboration of kinematics of knots in the radio jets in a sequence of images, a kinematic model of the radio jets with a velocity of $\approx 0.20\,c$ is a little better than one with the canonical velocity 0.26\,$c$ observed actually in the optical and X-ray jets of SS\,433. 
Consequently the distance to SS\,433 should be lowered to 4.3~kpc.

Besides this difference in velocity, the shift of the radio jets in the precession phase reveals itself in non-transient fashion, whereas it is not observed in the optical jets. 
In light of these differences, the jets must have two-component structure: with on-axis channel -- optical jets, and low velocity shell -- radio jets. 

\end{abstract}

\section*{\large The problem of the distance to SS\,433}
The distance determined from different methods based on observations of signatures of the diffuse interstellar medium is placed in the range from 2.2~kpc to 5.5~kpc, and it is disputable yet (e.g. \cite{Lock07}) because of big uncertainties inherent to these methods.
Therefore by widespread opinion the most rigorous estimate is a distance of 5.5~kpc as obtained from proper motion measurements of the radio jets of SS\,433, which accounts for the light travel effect.
This method allows to determine jet velocity $v_{\rm j}$ and the distance $D$ independently of each other.
However, this estimation is based on the observations whose accuracy is not adequate for determination of the distance. 

The distance derivation by Blundell \& Bowler (2004) is accepted currently as a corner stone in this question. 
Their estimation assumes that the twin jets are symmetrical perfectly, while observations do not show one to one correspondence between knots in both radio jets. 
Therefore the confirmation of the symmetry in the image in \cite{BB04}, by simulation of jet track on the sky plane, seems very strange.
Moreover, their results are contradictory: the positions of the colored points in their Fig.~3, which trace the ridgeline of the jets, do not correspond to the velocity distribution along the jets in their Fig.~4a. 
That is shown in Fig.~1 where we represent Fig.~3 from \cite{BB04} with the overlaid rings corresponding to the velocity distribution -- they deviate significantly from the ridgeline.
In that case the estimation $D=5.5$~kpc by Blundell \& Bowler (2004) can be affected by this discrepancy.

Here we revalue the distance to SS\,433 by applying the kinematic model to a sequence of the images of the radio jets presented by Stirling et al. (2002). 
A radio jet in this sequence is time resolved and thus the kinematic model is applicable to movement of discrete knots in the jet. 
The applicability to one image is not evident because various effects of continuous flow, which are quite possible in jets, could influence brightness distribution in the jet and knots kinematics as a result.

\section*{\large Kinematic model}
Kinematic model assumes the SS\,433 jets consist of independent blobs that move at a constant velocity, and determines direction of the jet velocity vector at a moment of ejection, which regularly precesses and nods (or nutates, e.g. see the scheme of the jet precession in Fig.~1 in \cite{CN86}).
The kinematic model is determined by following ephemerides:\\[0.3cm]
\vspace{0.3cm}
\begin{tabular}{p{10cm}|p{4.0cm}|p{0.5cm}}
 \hline
                                       &                          & ref.  \\
 \hline
 position angle of jet precession axis & $\chi = 98^\circ.2$      & \cite{St02} \\
 inclination of the axis to the line of sight & $i = 78^\circ.81$ & \cite{Dav08}\\
 precession cone half-angle            & $\theta_0 = 19^\circ.75$ & \cite{Dav08}\\
 precession period                     & $P_0 = 162.250$\,d       & \cite{Dav08}\\
 initial precession phase epoch (of minimal inclination of the east jet to the
 line of sight)                        & $t_0 =$\,JD\,2\,443\,508.41 & \cite{Dav08}\\ 
 orbital period of the system          & $P_{\rm b}= 13.08211$\,d & \cite{Gor98}\\
 amplitudes of sinusoidal oscillation of phase and cone opening of the precession 
 due to nutation                       & $\delta \psi_0 =2^\circ.9$, 
                                         $\delta \theta_0 =2^\circ.6$ & \cite{NC86}\\
 initial phases of these motions       & $\phi_{0\psi} =241^\circ$,
                                         $\phi_{0\theta} =104^\circ$  & \cite{NC86}\\
 epoch of these phases                 & $t_{\rm 0n}=$\,JD\,2\,443\,650.958 & \cite{NC86}\\
\hline
\end{tabular}

\noindent where $t_{\rm 0n}$ is the epoch of Newsom \& Collins (1986) corrected
by $0.958$\,d to satisfy the nutation phase reference time of 
\cite{Dav08} obtained from much bigger database than one of \cite{NC86}. 
In the frame of this kinematic model an outflow velocity of
$\sim 0.26\,c$ in the jets follows from Doppler shifts of the moving optical and X-ray lines.

The apparent proper motion of a blob in jets is determined by
\beq
\mu = \frac{v_\tau} {1+v_{\rm r}/c} \frac{1} {D},
\label{prop}
\eeq
where $v_\tau$ and $v_{\rm r}$ are the projections of the velocity vector
on the sky plane and on the line of sight as determined by the kinematic model, $c$ is the speed of light.
This formula accounts for the light travel effect, and it is applicable only in the case of uniform straightforward movement.
Otherwise, one has to use iteration technique of calculations.

\section*{\large Results of simulation of kinematics of the radio jets}
Free parameters of the simulation were the jet velocity $v_{\rm j}$ and the amplitude of the proper motion $\mu_{\rm m}= v_{\rm j}/D$ -- the parameters, which could be determined independently of each other {\it only} from the light travel effect in the radio jets; the shift of precession phase and the jet acceleration -- because they are suspected (e.g. \cite{St02,St04}).
Fig.~2 shows the jet tracks fitted to the knots on a sequence of the images from \cite{St02}. 
The comparison of the best model, in residuals with respect to the knots, to the canonical one, with the velocity of the optical jets, 0.258\,$c$ \cite{Dav08}, is presented in the following table:\\[0.3cm]
\vspace{0.3cm}
\begin{tabular}{l|c|c|c|c}
\hline
model & velocity & distance & phase shift& residuals per knot (\# of knots) \\
&$\beta=v_{\rm j}/c$&$D$ & $\Delta \psi \cdot P_0$&$\sigma$     \\
&            & (kpc)     & (day)        &($0''.001$) \\
\hline
 "the best" $^{\rm a}$& $0.205\pm 0.006$ & $4.36\pm 0.12$ & $7.8\pm 0.2$   & 6.93 (27)\\
\hline
 canonical  & 0.258 & 5.5 & 7.5   & 8.99 (27)\\
\hline
\end{tabular}

\noindent $\mbox{ }^{\rm a}$ the errors are due to the uncertainties of the nutation 
ephemeris \cite{NC86}, the main source of uncertainties of the kinematic model

1. The acceleration of the jets influence on the residuals of the model insignificantly.
Thus, from our simulation the acceleration is not discernible.
However, observed peculiar knots (e.g. see Fig.~2) and deviations of the radio jets from the kinematic model \cite{St02,Sch04} could be due to the acceleration, perhaps variable one.

2. The velocity $v_{\rm j}=0.205\,c$ gives a little better fit than
$v_{\rm j}=0.258\,c$. 
The difference in $\sigma$ between the models is small, $\sim 30\%$.
This difference corresponds to the standard deviation between the models
\mbox{$\sigma_{\rm l.t.e.} = \sqrt{9.04^2 - 6.99^2} = 5.73$\,mas} per knot.
That is anticipated from the light travel effect: due to the optimization of a model the residual per knot of another model fit changes in half a difference between 
two models $\mu_{\rm m} (\beta_2-\beta_1) \langle t |\sin(2\theta)| 
\rangle/2 = 9.0$\,mas on the average per knot, as derived from expression (\ref{prop}), where $\theta$ is the 
direction angle of the jet velocity vector to the line of sight, 
$t$ is the time of flight of knots in jets.
Thus, the positions of the knots in the jets are fitted in the best way by the model tracks for a distance of 4.3\,kpc.
(The similar work was done by Stirling et al. (2002), but for proper motions of the knots.)

3. The radio jets lead the kinematic model prediction in the precession phase, by $\sim 8$ days, that was discovered by Stirling et al. (2002) yet, and this lagging is reversed for the jet in the image of Blundell \& Bowler (2004, see our Fig.~1). 
These phase shifts take place for the whole radio jets in the images, i.e. for a time in order of the precession period, and do not depend on an accepted magnitude of the velocity in the radio jets.
That is not observed in the redshifts of the moving lines H$_\alpha$ in Fig.~3 obtained from the database of Iijima (1993) and Davydov, Esipov \& Cherepashchuk (2008).

All together, these facts indicate essential difference in kinematics between the optical and radio jets. 
This assumes the two-component structure of the jets of SS\,433, with a high-speed spin and low speed shell. 
Indeed, the similar transverse structure appears in high-resolution observations of AGN jets in the last years. 
Such structure might naturally explain observed small deviations of the radio jets of SS\,433 from the kinematic model by the local acceleration and deceleration of blobs in the jets under interaction of these components.

\newpage

{\Large \bf Figure captions}
\vspace{0.5cm}

Fig.~1. 
VLA map of the precessing radio jets of SS\,433 on 2003 July 10 (JD\,2\,452\,831) 
from \cite{BB04}; the horizontal and vertical axes are right ascension and 
declination (both in units of $0''.5$), respectively; the resolution at the 
ridgeline of the jet is 35~mas. 
The colored points, which trace the ridgeline, deviate by eye from the colored 
rings -- the jets regions, which the ejection times and velocities determined 
by Blundell \& Bowler (2004) for the colored points should correspond to, as 
it follows from the canonical kinematic model.
The helical lines are the model jet tracks with the velocities $v_{\rm j}$ of 
0.258\,$c$ and 0.205\,$c$ (dotted line) and the same $v_{\rm j}/D$ ratio of 
8.12~mas/day; the blue and red colors correspond respectively to the approaching 
and receding regions in the jets. 
It is seen that these radio jets (the colored points) lag the model track, the 
kinematic model prediction; the lagging is $\sim 9$~days.

Fig.~2. 
Contoured maps of radio jets of SS\,433 from \cite{St02}, from observations 
with MERLIN telescopes on 1991 December 7, 12 and 22, and on 1992 January 4 
(JD\,2\,448\,626.52), from up to bottom; the horizontal and vertical axes are 
right ascension (in units of $0''.2$) and declination (in units of $0''.1$), 
respectively. 
We fitted the simulated tracks on the sky plane to the bright knots in the 
images, that assumes the knots lie at projection of jet axis on the sky plane. 
The knots positions (the crosses) had been determined in \cite{St02} as 
centers of Gaussian profile; their accuracy is of order several mas \cite{Par09}. 
The circles mark the knots, which are used in the fitting procedure: there are 
excluded the inner knots (closer than $0''.2$), where the light travel effect 
is insignificant, and the knots with big deviations from the kinematic model. 
A consistency check shows very close coincidence
of the proper motions of the used knots with the kinematic model prescription.
The jets tracks in the images are obtained with account for the nutation and 
correspond to the best (solid line) and canonical (thin line) models.

Fig.~3. 
The redshifts of the moving lines H$_\alpha$ of the SS\,433 jets from the 
database of Davydov, Esipov \& Cherepashchuk (2008) (top) and  Iijima (1993) 
(bottom) in the time intervals spanning the formation times of the jets in Fig.1 
and 2, respectively. 
The model redshift curves were derived using the kinematic model with the 
velocity $v_{\rm j}=0.258\,c$, the blue (red) curve corresponds to the east 
(west) jet in the figures. 
It is worth noting that a nutation wave on the curves spans 6 days, and an 
apparent shift by around 8 days between the data and the curves is not seen. 
On the contrary, the data shows very good phasing with the curves, in spite 
of discrepancy in amplitude.

\newpage
%
\begin{figure}[ht]
\resizebox{\hsize}{!}{\includegraphics{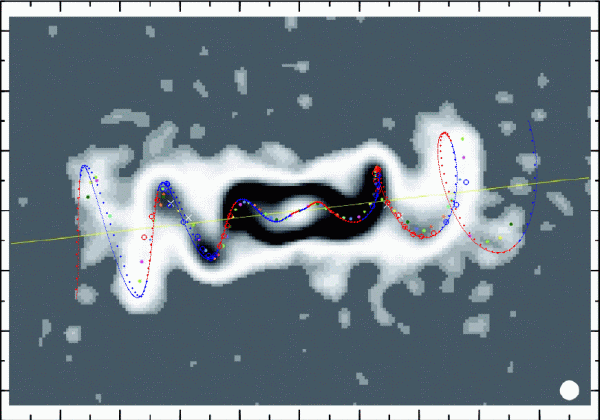}}
\caption{
VLA map of the precessing radio jets of SS\,433 on 2003 July 10 (JD\,2\,452\,831) 
from \cite{BB04}; the horizontal and vertical axes are right ascension and 
declination (both in units of $0''.5$), respectively; the resolution at the 
ridgeline of the jet is 35~mas. 
The colored points, which trace the ridgeline, deviate by eye from the colored 
rings -- the jets regions, which the ejection times and velocities determined 
by Blundell \& Bowler (2004) for the colored points should correspond to, as 
it follows from the canonical kinematic model.
The helical lines are the model jet tracks with the velocities $v_{\rm j}$ of 
0.258\,$c$ and 0.205\,$c$ (dotted line) and the same $v_{\rm j}/D$ ratio of 
8.12~mas/day; the blue and red colors correspond respectively to the approaching 
and receding regions in the jets. 
It is seen that these radio jets (the colored points) lag the model track, the 
kinematic model prediction; the lagging is $\sim 9$~days.
}
\end{figure}
%

\newpage
%
\begin{figure}[ht]
\begin{center}
\resizebox{!}{\vsize}{\includegraphics*{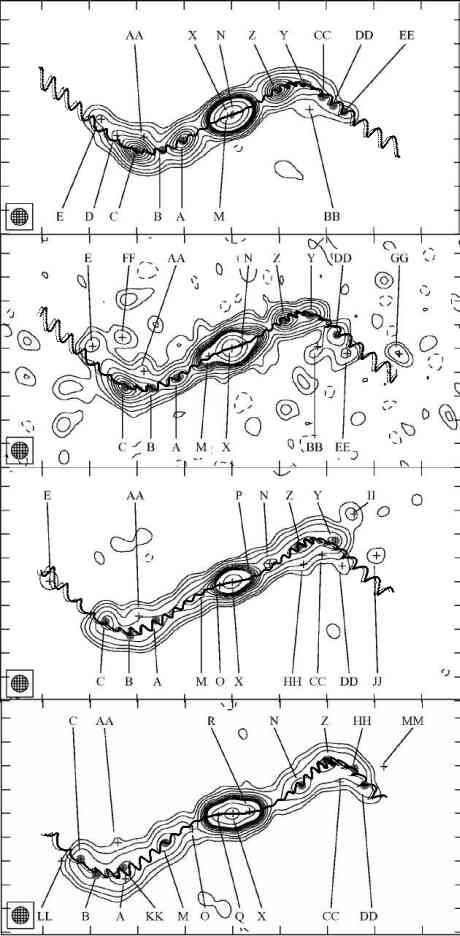}}
\end{center}
\caption[]{
}
\end{figure}
%

\newpage
%
\begin{figure}[ht]
\resizebox{!}{8cm}{\includegraphics*{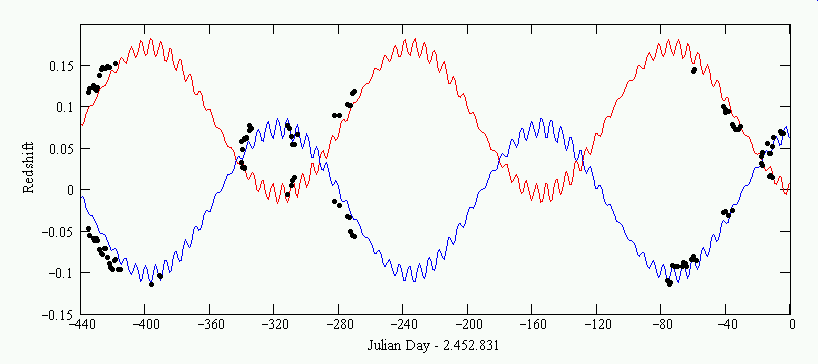}}

\resizebox{!}{8cm}{\includegraphics*{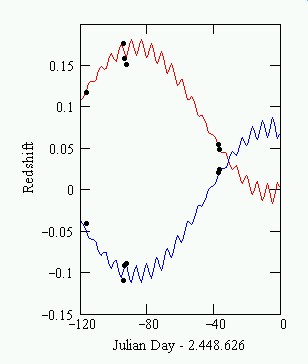}}
\caption[]{
The redshifts of the moving lines H$_\alpha$ of the SS\,433 jets from the 
database of Davydov, Esipov \& Cherepashchuk (2008) (top) and  Iijima (1993) 
(bottom) in the time intervals spanning the formation times of the jets in Fig.1 
and 2, respectively. 
The model redshift curves were derived using the kinematic model with the 
velocity $v_{\rm j}=0.258\,c$, the blue (red) curve corresponds to the east 
(west) jet in the figures. 
It is worth noting that a nutation wave on the curves spans 6 days, and an 
apparent shift by around 8 days between the data and the curves is not seen. 
On the contrary, the data shows very good phasing with the curves, in spite 
of discrepancy in amplitude.
}
\end{figure}
%

\end{document}